\documentclass[11pt]{article}

\usepackage{amsbsy}   
\usepackage{amsfonts}   
\usepackage{amsmath}    
\usepackage{amssymb}
\usepackage{slashed}
\usepackage{cite}
\usepackage{url}

\usepackage{psfrag}
\usepackage[latin1]{inputenc}
\usepackage[dvips]{graphicx} 
\usepackage{slashed}
\usepackage{mciteplus}
\usepackage{hyperref}
\usepackage{booktabs}

\newcommand{\sd}{\mathrm{d}}

\newcommand{\kd}{\kappa^{D}}
\newcommand{\rd}{\rho^{D}}
\newcommand{\ku}{\kappa^{U}}
\newcommand{\ru}{\rho^{U}}
\newcommand{\kl}{\kappa^{L}}
\newcommand{\rl}{\rho^{L}}
\newcommand{\kf}{\kappa^{F}}
\newcommand{\rf}{\rho^{F}}
\newcommand{\sinb}{\sin\beta}
\newcommand{\cosb}{\cos\beta}
\newcommand{\tb}{\tan\beta}
\newcommand{\ctb}{\cot\beta}
\newcommand{\sba}{s_{\beta-\alpha}}
\newcommand{\cba}{c_{\beta-\alpha}}

\newcommand{\mhref}{m_{h,\mathrm{ref}}}

\newcommand*{\Ave}[1]{\mathinner{\left\langle{#1}\right\rangle}}
\newcommand{\cp}{\mathcal{CP}}
\newcommand{\thdmc}{{\tt 2HDMC}}
\newcommand{\THDM}{{\tt THDM}}
\newcommand{\SM}{{\tt SM}}
\newcommand{\DecayTable}{{\tt DecayTable}}
\newcommand{\Constraints}{{\tt Constraints}}
\newcommand{\NMSSMT}{{\tt NMSSMTools}}
\newcommand{\HB}{{\tt HiggsBounds}}
\newcommand{\MGME}{{\tt MadGraph/MadEvent}}
\usepackage{url}

\begin{document}
\pagestyle{empty}

\begin{center}
\Large\bf\boldmath
\vspace*{0.8cm} {\tt 2HDMC} - Two-Higgs-Doublet Model Calculator
\unboldmath
\Large
\end{center}

\vspace{0.2cm}
\begin{center}
\large
David Eriksson\footnote{Electronic address: \tt david.eriksson@physics.uu.se}, Johan Rathsman\footnote{Electronic address: \tt johan.rathsman@physics.uu.se}, and Oscar St\aa l\footnote{Electronic address: \tt oscar.stal@physics.uu.se} \\
\vspace{0.4cm}
{\sl High-Energy Physics, Dept.~of Physics and Astronomy,\\ Uppsala University, P.\,O.\,Box 516, SE-751\,20 Uppsala, Sweden}\\
\vspace{0.4cm}
\large
September 18, 2009\\
\end{center}
\vspace{0.6cm}

\begin{abstract}
\noindent 
This manual describes the public code \thdmc\ which can be used to perform calculations in a general, $\cp$-conserving, two-Higgs-doublet model (2HDM). The program features simple conversion between different parametrizations of the 2HDM potential, a flexible Yukawa sector specification with choices of different $Z_2$-symmetries or more general couplings, a tree-level decay library including all two-body -- and some three-body -- decay modes for the Higgs bosons, and the possibility to calculate observables of interest for constraining the 2HDM parameter space, as well as theoretical constraints from positivity and unitarity.\\
\\
The latest version of the \thdmc\ code and full documentation is available from:\\ \url{http://www.isv.uu.se/thep/MC/2HDMC}
\end{abstract}

\newpage
$\quad$
\newpage
\tableofcontents
\newpage
\pagestyle{plain}
\setcounter{page}{1}

\section{Introduction}
In many respects, the two-Higgs-doublet model (2HDM) provides the simplest extension of the Standard  Model (SM) Higgs sector. For a general introduction to the model the reader is referred to \cite{Gunion:1989we}, or to the more recent papers \cite{Gunion:2002zf,Ginzburg:2004vp,Davidson:2005cw,*Haber:2006ue, Diaz:2002tp,*Gunion:2005ja,*Maniatis:2006fs,*Ivanov:2006yq,*WahabElKaffas:2007xd,*Ginzburg:2007jn,*Gerard:2007kn,*Maniatis:2007vn,*Ivanov:2007de,*Sokolowska:2008bt,*Maniatis:2009vp,*Ferreira:2009wh,*deVisscher:2009zb} dealing with various aspects of the 2HDM. The extensive review \cite{Djouadi:2005gj}, discussing the role of the 2HDM in the MSSM, might also be useful.

Despite its simplicity, the general 2HDM is an interesting theoretical laboratory, adding possible new phenomena such as spontaneous and explicit $\cp$-violation, lepton number violation, charged Higgs bosons, a dark matter candidate, and a heavy neutral Higgs boson compatible with electroweak precision data. Since the 2HDM suffers the same quadratic divergences in corrections to the scalar masses as the SM, these precise theoretical constructs of the 2HDM are perhaps not likely to be manifested in a more complete theory of nature -- and certainly not all of them at the same time. However, the large theoretical interest in these models calls for the development of phenomenological tools to help theorists provide experimentalists with viable models. This is the purpose of the two-Higgs-doublet model calculator (\thdmc) which we describe here.

\thdmc\ (Two-Higgs-Doublet Model Calculator) is a {\tt C++} code which can be used to study the phenomenology of a general 2HDM. At the moment, only the $\cp$-conserving 2HDM has been implemented. The Higgs potential in \thdmc\ can be specified in different parametrizations, allowing for everyone to use their own favorite. The Yukawa sector can be specified with full freedom -- including specifying couplings which lead to flavour-changing neutral currents. Some standard choices (like type I, type II etc.) can also be used. The code provides methods to check theoretical properties of the model, such as positivity of the potential and $S$-matrix unitarity. There are also methods for calculating the 2HDM contribution to the oblique electroweak parameters $(S,T,U,V,W,X)$, to determine the contribution to $(g-2)_\mu$, and to check the model against constraints from colliders. In addition, decay widths of the Higgs bosons can be calculated.
The whole code is fully modular and based on object-oriented principles. It can be used as a library, in stand-alone applications, or exchanging data with other codes through a LesHouches-inspired format.

The organization of this paper is as follows: section~\ref{sect:physics} contains a short, hopefully self-contained, overview of the physics of the 2HDM. The focus here lies, of course, on the aspects relevant for \thdmc. Section~\ref{sect:constraints} discusses theoretical and experimental ways of constraining the 2HDM parameter space, while section~\ref{sect:code} describes the structure of the \thdmc\ code itself and how to use it. How to obtain and compile the code is described in section~\ref{sect:install}. If you are already familiar with the 2HDM and want to get started right away, you should probably skip straight to section~\ref{sect:code}.

\section{Physics of the 2HDM}
\label{sect:physics}
\subsection{The Higgs potential}
Denote a complex scalar $SU(2)_{\rm{L}}$ doublet field $\Phi$. Assigning a hypercharge $Y=1$, the component representation of $\Phi$ becomes $\Phi(x)=(\Phi^+(x), \Phi^0(x))^T$. In its most general form, the 2HDM contains two identical such fields $\Phi_{1}$ and $\Phi_2$, related by a global $U(2)$ symmetry under which $\Phi_a\to U_{ab}\Phi_b$ and $\Phi_a^{\dagger}\to \Phi_b^{\dagger}U_{ba}^{\dagger}$.
In the conventions of \cite{Davidson:2005cw,*Haber:2006ue}, the most general gauge invariant and renormalizable potential that can be formed using two Higgs doublets is given by
\begin{equation}
  \begin{aligned}
    \mathcal{V_{\rm{gen}}} = &m_{11}^2\Phi_1^\dagger\Phi_1+m_{22}^2\Phi_2^\dagger\Phi_2
    -\left[m_{12}^2\Phi_1^\dagger\Phi_2+\mathrm{h.c.}\right]
    \\
    &+\frac{1}{2}\lambda_1\left(\Phi_1^\dagger\Phi_1\right)^2
    +\frac{1}{2}\lambda_2\left(\Phi_2^\dagger\Phi_2\right)^2
    +\lambda_3\left(\Phi_1^\dagger\Phi_1\right)\left(\Phi_2^\dagger\Phi_2\right)
    +\lambda_4\left(\Phi_1^\dagger\Phi_2\right)\left(\Phi_2^\dagger\Phi_1\right)
    \\&+\left\{
    \frac{1}{2}\lambda_5\left(\Phi_1^\dagger\Phi_2\right)^2
    +\left[\lambda_6\left(\Phi_1^\dagger\Phi_1\right)
      +\lambda_7\left(\Phi_2^\dagger\Phi_2\right)
      \right]\left(\Phi_1^\dagger\Phi_2\right)
    +\mathrm{h.c.}\right\}.
  \end{aligned}
  \label{eq:pot_gen}
\end{equation}
The parameters $m_{11}^2,m_{22}^2$ and $\lambda_{1-4}$ are real numbers (since the potential has to be real), whereas the remaining parameters $\lambda_{5-7}$ and $m_{12}^2$ in general can be complex. Non-zero imaginary parts of the complex parameters which cannot be removed by a rephasing transformation give rise to explicit $\cp$-violation in the Higgs sector. Since we are not going to treat $\cp$-violating effects here, we assume all parameters to be real in the following. 

Since the potential given by eq.~(\ref{eq:pot_gen}) is manifestly $U(2)$-invariant, values specified for the parameters $\{m_{ij}^2, \lambda_i\}$ can only have definite (physical) meaning when a particular basis is specified for the scalar fields. Alternatively, one may reformulate the 2HDM using a fully basis-invariant language \cite{Davidson:2005cw,*Haber:2006ue} (see also \cite{Ivanov:2005hg} for a group theoretic analysis of the basis invariance). Since this formalism has so far been sparingly applied in 2HDM phenomenology, we will maintain the notion of choosing a particular basis. Expressed in terms of vacuum expectation values (vevs), a generic basis respecting the $U(1)_{\rm{EM}}$ gauge symmetry can be written as
\[
\Ave{\Phi_1} = \frac{v}{\sqrt{2}}\left(
\begin{array}{c}
0 \\
\cos\beta
\end{array}
\right)\quad\quad
\Ave{\Phi_2} = \frac{v}{\sqrt{2}}\left(
\begin{array}{c}
0 \\
e^{i\xi}\sin\beta 
\end{array}
\right),
\]
where $v=(\sqrt{2}G_F)^{-1/2}\approx246$ GeV. By convention $0\leq \beta \leq \pi/2$ is chosen. A non-zero phase $\xi$ results in a vacuum which breaks $\cp$ spontaneously. Since we do not aim to discuss $\cp$-violating effects, we take $\xi=0$.
Note that the value of $\tan\beta\equiv \Ave{\Phi_2}/\Ave{\Phi_1}$ at this point only determines one particular choice of basis. Since the 2HDM potential is invariant under a change of this basis, $\tan\beta$ can not be a physical parameter of the model in general.
Within the set of $\cp$-conserving bases defined above there exists a special choice, called the \emph{Higgs basis}, in which only one of the two doublets is assigned a non-zero vev.

\subsection{Electroweak symmetry breaking}
The electroweak $SU(2)\times U(1)$ symmetry manifest in eq.~(\ref{eq:pot_gen}) is spontaneously broken down to $U(1)_{\mathrm{EM}}$ by a negative eigenvalue of the scalar mass matrix $m_{ij}^2$, causing at least one of the Higgs doublets to develop a vev. In a basis where $0<\tan\beta< \infty$, the two minimization conditions
\begin{equation}
m_{11}^2 = m_{12}^2\tan\beta-\frac{1}{2}v^2\Bigl(\lambda_1\cos^2\beta+\left(\lambda_3+\lambda_4+\lambda_5\right)\sin^2\beta+3\lambda_6\sin\beta\cos\beta+\lambda_7\sin^2\beta\tan\beta\Bigr),
\label{eq:m11}
\end{equation}
\begin{equation}
m_{22}^2 = m_{12}^2\cot\beta-\frac{1}{2}v^2\Bigl(\lambda_2\sin^2\beta+\left(\lambda_3+\lambda_4+\lambda_5\right)\cos^2\beta+\lambda_6\cos^2\beta\cot\beta+3\lambda_7\sin\beta\cos\beta\Bigr)
\label{eq:m22}
\end{equation}
can be used to eliminate $m_{11}^2$ and $m_{22}^2$ from eq.~(\ref{eq:pot_gen}). Similar equations for the minimum conditions, not given here, exist for the Higgs basis. Eliminating $m_{11}^2$ and $m_{22}^2$ leaves eight real parameters in the Higgs potential, not counting $\tan\beta$ and $v$.

When the symmetry is broken, the eight degrees of freedom from the $SU(2)$ doublets $\Phi_1$ and $\Phi_2$ are usually re-expressed in states with definite physical properties. The spectrum then contains three Goldstone modes: $G^\pm$ and $G^0$, which are absorbed to give mass to the gauge bosons $W^\pm$ and $Z$, reducing the number of physical Higgs states to five. Three of these states are neutral, of which two ($h$ and $H$, with $m_h \leq m_H$) are $\cp$-even, and one (typically denoted $A$) is $\cp$-odd. The remaining two are a pair of charged Higgs bosons ($H^\pm$). The charged Higgs bosons are a characteristic of the 2HDM not present in the SM.

Introducing the mixing angle $\alpha$ to diagonalize the mass matrix for the $\cp$-even states, the original doublets are expressed as
\begin{equation}
\Phi_1=\frac{1}{\sqrt{2}}\left(\begin{array}{c}
\displaystyle \sqrt{2}\left(G^+\cos\beta -H^+\sin\beta\right)  \\
\displaystyle v\cos\beta-h\sin\alpha+H\cos\alpha+\mathrm{i}\left( G^0\cos\beta-A\sin\beta \right)
\end{array}
\right)
\end{equation}
\begin{equation}
\Phi_2=\frac{1}{\sqrt{2}}\left(\begin{array}{c}
\displaystyle \sqrt{2}\left(G^+\sin\beta +H^+\cos\beta\right)  \\
\displaystyle v\sin\beta+h\cos\alpha+H\sin\alpha+\mathrm{i}\left( G^0\sin\beta+A\cos\beta \right)
\end{array}
\right).
\end{equation}
The Higgs masses can now easily be calculated for any choice of Higgs potential and basis. Using the generic potential of eq.~(\ref{eq:pot_gen}), one arrives at the relations \cite{Gunion:2002zf}
\begin{equation}
\begin{aligned}
m_A^2&=\frac{m_{12}^2}{\sinb\cosb}-\frac{v^2}{2}\left(2\lambda_5+\lambda_6\ctb+\lambda_7\tb\right)\\
m_{H^+}^2&=m_A^2+\frac{v^2}{2}\left(\lambda_5-\lambda_4\right).
\end{aligned}
\end{equation}
Since the $\cp$-even states mix, their mass matrix $\mathcal{M}$ is given by
\begin{equation}
\mathcal{M}^2=m_A^2\left(\begin{array}{cc}
s^2_\beta & -s_\beta c_\beta \\
-s_\beta c_\beta & c^2_\beta
\end{array}
\right)
+v^2 \mathcal{B}^2,
\end{equation}
where 
\begin{equation}
\mathcal{B}^2=\left(\begin{array}{cc}
\lambda_1 c^2_\beta+2\lambda_6s_\beta c_\beta+\lambda_5 s^2_\beta & 
\left(\lambda_3+\lambda_4\right)s_\beta c_\beta+\lambda_6 c^2_\beta+\lambda_7 s^2_\beta \\
\left(\lambda_3+\lambda_4\right)s_\beta c_\beta+\lambda_6 c^2_\beta+\lambda_7 s^2_\beta & 
\lambda_2 s^2_\beta+2\lambda_7s_\beta c_\beta+\lambda_5 c^2_\beta 
\end{array}
\right).
\end{equation}
Performing a rotation by $\alpha$ to diagonalize $\mathcal{M}$, one obtains the masses of the $\cp$-even states
\begin{equation}
\left(\begin{array}{cc}
m_H^2 & 0 \\
0 & m_h^2
\end{array}
\right)=\mathcal{R}(\alpha)\mathcal{M}^2\mathcal{R}^T(\alpha)
\end{equation}
with $m_H^2\geq m_h^2$. Explicitly, the mass eigenvalues are given by \cite{Gunion:2002zf}
\begin{equation}
m_{H,h}^2=\frac{1}{2}\left[\mathcal{M}_{11}^2+\mathcal{M}_{22}^2\pm\sqrt{\left(\mathcal{M}_{11}^2-\mathcal{M}_{22}^2\right)^2+4\left(\mathcal{M}_{12}^2\right)^2}\;\right].
\end{equation}

Because of gauge invariance, and the way masses has to be supplied to the EW gauge bosons, all couplings of Higgs bosons to gauge bosons are completely determined in terms of the invariants $\sba\equiv \sin(\beta-\alpha)$ and $\cba\equiv \cos(\beta-\alpha)$. For this important angle we find it convenient to use a convention in which $-\pi/2 \leq \beta-\alpha \leq \pi/2$. This choice allows the basis to be changed by varying $\tan\beta$ while preserving the invariants $\sba$ and $\cba$.
 It is important to respect this convention when working explicitly with the angle $\alpha$, e.g.~when interfacing \thdmc\ with other codes. 

\subsection{The Yukawa sector}
In the following we will assume the same fermion content as in the Standard Model (with massless neutrinos). The mass eigenstates for the down- and up-type quarks are written as vectors in flavour space, denoted $D$ and $U$ respectively, and similarly for the leptons $L$ and $\nu$. The 2HDM Yukawa sector, expressed in the physical Higgs mass eigenstates $(h,H,A,H^\pm)$, is then given in basis-independent form as \cite{Davidson:2005cw,*Haber:2006ue}
\begin{equation}
  \begin{aligned}
    -\mathcal{L}_{\rm{Yukawa}}=&\frac{1}{\sqrt{2}}\overline{D}\Bigl\{\kd\sba+\rd\cba \Bigr\}Dh
    +\frac{1}{\sqrt{2}}\overline{D}\Bigl\{\kd\cba-\rd\sba \Bigr\}DH+ \frac{\mathrm{i}}{\sqrt{2}}\overline{D}\gamma_5\rd DA \\
    &+\frac{1}{\sqrt{2}}\overline{U}\Bigl\{\ku\sba+\ru\cba \Bigr\}Uh
    +\frac{1}{\sqrt{2}}\overline{U}\Bigl\{\ku\cba-\ru\sba \Bigr\}UH- \frac{\mathrm{i}}{\sqrt{2}}\overline{U}\gamma_5\ru UA \\
    &+\frac{1}{\sqrt{2}}\overline{L}\Bigl\{\kl\sba+\rl\cba \Bigr\}Lh
    +\frac{1}{\sqrt{2}}\overline{L}\Bigl\{\kl\cba-\rl\sba \Bigr\}LH+ \frac{\mathrm{i}}{\sqrt{2}}\overline{L}\gamma_5\rl LA \\
    &+\Bigl[\overline{U}\bigl\{V_{\rm{CKM}}\rd P_R-\ru V_{\rm{CKM}} P_L\bigr\}DH^+ + \overline{\nu}\rl P_RL H^+ + \rm{h.c.}\Bigr].  
\end{aligned}
\label{eq:yukawa}
\end{equation}
The diagonal $3\times 3$ matrices $\kappa^F$ are given by $\kappa^F\equiv\sqrt{2}M^F/v$, where $M^F$ are the corresponding mass matrices for the fermions ($F=D,U,L$), and as usual $P_{R/L}=(1\pm\gamma_5)/2$. The generality of this Yukawa Lagrangian is present through the freedom to choose the components of each $\rho^F$ arbitrarily. A symmetric $\rho^F$, as required by $\cp$ conservation, has six independent components. In practice, the allowed size of the off-diagonal elements in $\rho^F$ are strongly constrained, since non-zero elements induce Higgs-mediated flavour-changing neutral currents (FCNC) at tree-level. Dealing with this problem through a careful adjustment of individual elements in $\rho^F$ involves severe fine-tuning. From a theoretical standpoint, this can therefore be deemed an unnatural solution.

To address this problem in a more elegant way, one can invoke the theorem of Glashow and Weinberg \cite{Glashow:1976nt}, stating that tree-level FCNC are absent in any theory where a given fermion does not couple to more than one Higgs doublet. This situation can be arranged by imposing a suitable $Z_2$ symmetry to distinguish $\Phi_1$ from $\Phi_2$. Assigning definite $Z_2$ quantum numbers also to the right-handed fermions, the undesired couplings can be avoided. When all fermions of a given type (up-quarks, down-quarks and leptons) have one common $Z_2$ quantum number, the combinatorial possibilities for the Yukawa couplings are enumerated in 2HDM language as types I, II, III, and IV \cite{Barger:1989fj}.\footnote{Type III is often used in different context referring to a more general class of models with a broken (e.g. by SUSY loop effects) $Z_2$ symmetry in the Yukawa sector. In such models $\tan\beta$ is \emph{not} a physical parameter.} If the $Z_2$ symmetry is to be explicit in the Yukawa sector, a dependence is introduced on the basis chosen for the Higgs sector. This way $\tan\beta$ is promoted to a physical parameter through its appearance in the Yukawa couplings. In table \ref{tab:types} we list the relations between $\rf$ and $\kf$ for the physically distinct types of Yukawa couplings realized with a $Z_2$ symmetry. We note finally that some of the terms present in the Higgs potential violate this symmetry. The $\lambda_6$ and $\lambda_7$ terms in particular lead to hard $Z_2$ violation and should therefore be suppressed, while an $m_{12}^2$ term in many cases can be tolerated since it only violates $Z_2$ softly (meaning that the symmetry is restored in the UV). This issue is discussed in some detail in \cite{Ginzburg:2004vp}.
\begin{table}
\centering
        
\begin{tabular}{cccccc}
   \toprule                
  & \multicolumn{4}{c}{Type}   \\
         & I & II & III & IV \\
         \midrule
         $\rd$ & $\kd\ctb$  & $-\kd\tb$ & $-\kd\tb$ & $\kd\ctb $\\
         $\ru$ & $\ku\ctb$  & $\ku\ctb$ & $\ku\ctb$ & $\ku\ctb $\\
         $\rl$ & $\kl\ctb$  & $-\kl\tb$  & $\kl\ctb$ & $-\kl\tb $\\
         \bottomrule
\end{tabular}
 \caption{Relations between Yukawa coupling matrices $\rf$ and $\kf$ in the four different types of 2HDM Yukawa sectors where all fermions of a given type $(F=U,D,L)$ share the same $Z_2$ quantum number.} 
\label{tab:types}
\end{table}

\subsection{Decay widths}
From the masses and couplings, decay widths of the Higgs bosons ($H_1=h$, $H_2=H$, $H_3=A$, $H_4=H^\pm$) into various modes are calculated. At the tree-level, the Higgs bosons may decay into pairs of fermions, pairs of gauge bosons, one gauge boson and another Higgs boson, or into pairs of lighter Higgs bosons. The decays $H_i\to\gamma\gamma$ and $H_i\to g g$ ($i=1\dots 3$) are induced at the one-loop level.

\subsubsection{$H_i\to f\bar{f}'$}
We denote the coupling of Higgs boson $H_i$ ($i=1\dots 4$) to a pair of fermions by $C_S$ for the scalar part, and by $C_P$ for the pseudoscalar part. As an example, we get from eq.~(\ref{eq:yukawa}) that the couplings of $hd\bar{d}$ are $C_S=\frac{\mathrm{1}}{\sqrt{2}}\left(\kd\sba+\rd\cba\right)$ and $C_P=0$. Only for the charged Higgs are both $C_S$ and $C_P$ (potentially) non-zero at the same time. Introducing kinematic variables $x_1=m_f/m_{H_i}$, and $x_2=m_{\bar{f}'}/m_{H_i}$, the decay width to fermions is given by
\begin{equation}
\displaystyle \Gamma(H_i\to f \bar{f}')=\frac{N_c m_{H_i}}{8\pi}\Bigl\{\bigl[1-\left(x_1+x_2\right)^2\bigr]|C_{S}|^2+\bigl[1-\left(x_1-x_2\right)^2\bigr]|C_{P}|^2\Bigr\}\lambda^{1/2}\left(1,x_1^2,x_2^2\right),
\label{eq:Hff}
\end{equation}
where as usual
\begin{equation}
\lambda(1,x,y)=\left(1-x-y\right)^2-4xy.
\label{eq:lambda}
\end{equation}
In eq.~(\ref{eq:Hff}), $N_c$ refers to the number of ``colour'' degrees of freedom for the given fermion, i.e. $N_c=3$ for quarks and $N_c=1$ for leptons. For the decay of a neutral Higgs into quarks, the QCD radiative corrections are known at order $\alpha_s^2$ \cite{Braaten:1980yq,*Drees:1990dq,*Gorishnii:1990zu}. When these are included, the corrected width reads
\begin{equation}
\Gamma = \Gamma_0\left[1+5.67\frac{\alpha_s\left(m_{H_i}\right)}{\pi}+(35.94-1.36n_f)\left(\frac{\alpha_s\left(m_{H_i}\right)}{\pi}\right)^2\right]
\label{eq:gamma_hff_nlo}
\end{equation} 
in the $\overline{\mathrm{MS}}$ scheme with $n_f$ light flavours. The $\alpha_s$ correction is applied also to the decay of the charged Higgs. For both neutral and charged Higgs decays, resummation of the leading logarithmic corrections is implemented to all orders by using the running $\overline{\mathrm{MS}}$ fermion masses $m_f(\mu)$ in the Higgs couplings. The renormalization scale is then set at the Higgs mass scale $\mu=m_{H_i}$.

\subsubsection{$H_i\to VV$}
With $k=m_V^2/m_{H_i}^2$, the on-shell decay width of a neutral Higgs boson $H_i$ ($i=1\dots 3$) to a pair of massive gauge bosons is given by
\begin{equation}
\Gamma(H_i\to VV)=\delta_V\frac{|C_{H_{i}VV}|^2\, m_{H_i}^3}{128\pi\, m_V^4}\left(1-4k+12k^2\right)\sqrt{1-4k}.
\end{equation}
The fact that two $Z$ bosons in the final state are identical is handled by the factor $\delta_V=2\,(1)$ for $V=W\,(Z)$. Note that the coupling $C_{H_iVV}$ in our convention has the dimension of mass, which makes the overall dimension of this expression check out. When $m_{H_i}<2 m_V$, the three-body decay with one off-shell gauge boson decaying into a pair of fermions can instead be considered. Following \cite{Rizzo:1980gz,*Keung:1984hn}, we express the Dalitz density in the scaled fermion energies $x_{1,2}=2E_{1,2}/m_{H_i}$ as
\begin{equation}
\frac{\sd\Gamma}{\sd x_1\sd x_2}(H_i\to VV^*\to V f_1\bar{f}_2)=K_{H_iVV}\frac{(1-x_1)(1-x_2)+k(2x_1+2x_2-3+2k)}{(1-x_1-x_2)^2+k\gamma_V},
\label{eq:Dalitz_HVV}
\end{equation}
where the scaled width is $\gamma_V=\Gamma_V^2/m_{H_i}^2$. Integrating between the kinematic boundaries
\begin{equation}
\begin{aligned}
1-x_2-k &< x_1 < 1-\frac{k}{1-x_2} \\
0 &< x_2 < 1-k
\end{aligned}
\label{eq:bounds}
\end{equation}
the off-shell width is obtained. The normalization factor is given by 
\begin{equation}
K_{H_iVV}=\frac{9\, G_F}{64\sqrt{2}\,\pi^3}|C_{H_iVV}|^2m_{H_i}\delta'_V,
\end{equation}
with $\delta'_W=1$ for the $W$ coupling to fermions, and 
\begin{equation}
\delta'_Z=\frac{7}{12}-\frac{10}{9}\sin^2\theta_W+\frac{40}{27}\sin^4\theta_W.
\label{eq:deltaZ}
\end{equation}

\subsubsection{$H_i\to VH_j$}
It is also possible to have the Higgs $H_i$ ($i=1\dots 4$) decay to a combination of one Higgs- ($j=1\dots 4$) and one massive gauge boson. The width in this case becomes
\begin{equation}
\Gamma(H_i\to VH_j)=\frac{|C_{H_{i}VH_{j}}|^2\, m_V^2}{16\pi^2\, m_{H_i}}\lambda\left(1,\frac{m_{H_i}^2}{m_{V}^2},\frac{m_{H_j}^2}{m_{V}^2}\right)\lambda^{1/2}\left(1,\frac{m_V^2}{m_{H_i}^2},\frac{m_{H_j}^2}{m_{H_i}^2}\right),
\label{eq:hvh}
\end{equation}
where $V$ denotes one of the gauge bosons $Z$, $W^+$, or $W^-$, such that the decays $H_i \to W^-H^+$ and 
$H_i \to W^+H^-$ are distinct.
In addition to the on-shell decay, it is interesting to consider the decay when the gauge boson is off-shell. Analogous to eq.~(\ref{eq:Dalitz_HVV}), the Dalitz density in the rescaled energies $x_1$ and $x_2$ of the final state fermions can be expressed \cite{Djouadi:1995gv} as
\begin{equation}
\frac{\sd\Gamma}{\sd x_1 \sd x_2}(H_i\to H_jV^*\to H_j f_1\bar{f}_2)=K_{HVH}\frac{(1-x_1)(1-x_2)-k_H}{(1-x_1-x_2-k_H+k_V)^2+k_V\gamma_V}.
\end{equation}
Here $k_H=m_{H_j}^2/m_{H_i}^2$, $k_V=m_V^2/m_{H_i}^2$, and $\gamma_V=\Gamma_V^2/m_{H_i}^2$. The integration boundaries are given by eq.~(\ref{eq:bounds}), with $k$ replaced by $k_V$. The overall normalization factor is
\begin{equation}
K_{HVH}=\frac{9\,G_F}{16\sqrt{2}\pi^3}|C_{H_{i}VH_{j}}|^2m_V^2\delta_V'
\end{equation}
where $\delta'_W=1$ and $\delta'_Z$ is given by eq.~(\ref{eq:deltaZ}).

\subsubsection{$H_i\to H_j H_k$}
For the decay to a pair of scalars ($i=1\dots 3$, $j,k=1\dots 4$) we only consider the on-shell width, since the three-body decay width depends on the subsequent decay modes of the Higgs bosons. With $x_j=m_{H_j}^2/m_{H_i}^2$, and $x_k=m_{H_k}^2/m_{H_i}^2$, the width is given by
\begin{equation}
\Gamma(H_i\to H_j H_k)=\left(2-\delta_{jk}\right)\frac{|C_{H_iH_jH_k}|^2}{32\pi}M_{H_i}\lambda^{1/2}\left( 1,x_j,x_k\right).
\end{equation}
Note that $\delta_{jk}=0$ is used for the decay into a pair of charged Higgs bosons.

\subsubsection{$H_i\to \gamma\gamma$}
The two-photon decay of the neutral Higgs bosons $H_i$ ($i=1\dots 3$) proceeds through a loop with contributions from fermions, $W$ bosons, and charged Higgs bosons. The width is known \cite{Ellis:1975ap, Spira:1995rr}, and can be summarized in the expression
\begin{equation}
\Gamma(H_i\to\gamma\gamma)=\frac{\alpha^2M_{H_i}^3}{256\pi^3 v^2}\Bigl(\left|S^\gamma(M_{H_i})\right|^2+\left|P^\gamma(M_{H_i})\right|^2\Bigr),
\end{equation}
where the form factors are
\begin{equation}
S^\gamma(M_{H_i})=2\sum_f N_c Q_f^2 C^S_{H_if\bar{f}} \frac{v}{m_f}F_s(\tau_f)-C_{H_iWW}\frac{v}{2m_W^2}F_1(\tau_W)-C_{H_iH^+H^-}\frac{v}{2m_{H^+}^2}F_0(\tau_{H^+})
\end{equation}
for the scalar amplitude, and
\begin{equation}
P^\gamma(M_{H_i})=2\sum_f N_c Q_f^2 C^P_{H_if\bar{f}} \frac{v}{m_f}F_p(\tau_f)
\end{equation}
for the pseudoscalar part. As usual, $N_c=3\,(1)$ for quarks (leptons), and $Q_f$ is the electric charge in units of the elementary charge $|e|$. The various functions appearing in these form factors are related in terms of a common scaling function $f(\tau)$:
\begin{equation}
\begin{aligned}
F_s(\tau)&=\tau^{-1}\Bigl[1+\left(1-\tau^{-1}\right)f(\tau)\Bigr]\qquad F_p(\tau)=\frac{f(\tau)}{\tau}\\
F_0(\tau)&=\tau^{-1}\left[\tau^{-1}f(\tau)-1\right]\qquad\qquad\;\; F_1(\tau)=2+3\tau^{-1}+3\tau^{-1}(2-\tau^{-1})f(\tau)
\end{aligned}.
\label{eq:Fs}
\end{equation}
The argument is $\tau_f=m_{H_i}^2/4m_f^2$ for the fermions and equivalently for the $W^\pm$ and $H^\pm$ contributions. Finally, the (complex) function $f(\tau)$ itself is given by the integral
\begin{equation}
f(\tau)=-\frac{1}{2}\int_0^1\frac{\sd y}{y}\ln\left[1-4\tau y(1-y)\right]=\left\{
\begin{array}{lr}
\arcsin^2\left(\sqrt{\tau}\right) & \tau \leq 1 \\
\displaystyle -\frac{1}{4}\left[\ln\left(\frac{\sqrt{\tau}+\sqrt{\tau-1}}{\sqrt{\tau}-\sqrt{\tau-1}}\right)-\mathrm{i}\pi\right]^2 & \tau > 1.
\end{array}
\right.
\label{eq:ftau}
\end{equation}
\subsubsection{$H_i\to gg$}
The two-gluon decay mode is handled in a way very similar to $H_i\to\gamma\gamma$. The width is \cite{Spira:1995rr}
\begin{equation}
\Gamma(H_i\to gg)=\frac{\alpha_s^2M_{H_i}^3}{32\pi^3 v^2}\Bigl(K_S^g\left|S^g(M_{H_i})\right|^2+K_P^g\left|P^g(M_{H_i})\right|^2\Bigr),
\end{equation}
with the form factors
\begin{equation}
S^g(M_{H_i})=\sum_q C^S_{H_iq\bar{q}} \frac{v}{m_q}F_s(\tau_q), \qquad
P^g(M_{H_i})=\sum_q C^P_{H_iq\bar{q}} \frac{v}{m_q}F_p(\tau_q)
\end{equation}
and the same expressions given by eqs.~(\ref{eq:Fs}), (\ref{eq:ftau}) for the loop functions. The kinematic variabe $\tau_q=m_{H_i}^2/4m_q^2$. In the expression for $\Gamma(H_i\to gg)$, the leading-order QCD corrections are known in the heavy quark limit ($m_q \gg M_{H_i}$). These are given by \cite{Spira:1995rr}
\begin{equation}
\begin{aligned}
K^g_S&=1+\frac{\alpha_s(M_{H_i}^2)}{\pi}\left(\frac{95}{4}-\frac{7}{6}n_f\right)\\
K^g_P&=1+\frac{\alpha_s(M_{H_i}^2)}{\pi}\left(\frac{97}{4}-\frac{7}{6}n_f\right),
\end{aligned}
\end{equation} 
where $n_f$ is the number of quark flavours for which $m_q<m_{H_i}$. As a word of caution, it should be said that the complete NLO corrections may deviate from what is given here by order $10\%$ or more when moving away from the parameter space region where the heavy quark limit is applicable, e.g. at high $\tan\beta$ when $A$ couples preferentially to $b$ quarks.

\subsection{Example 1: Tree-level MSSM}
\label{sect:MSSM}
As is well known, the supersymmetric version of the 2HDM requires the Yukawa couplings of the model to be of type II. Opposite hypercharges for the Higgs doublets, acting as $Z_2$ quantum numbers, are introduced by the requirement that the superpotential is analytic. In addition, most of the parameters in the potential are fixed by supersymmetry to values given by the electromagnetic and weak gauge couplings. As a result the Higgs sector of the MSSM can be completely described, at tree-level, by the two inputs $m_A$ and $\tan\beta$. For completeness we give here the potential parameters of eq.~(\ref{eq:pot_gen}) expressed in the electroweak gauge couplings $g$ and $g'$ \cite{Inoue:1982pi,*Gunion:1984yn,*Carena:2002es}:
\[
\begin{aligned}
\lambda_1=\lambda_2=\frac{g^2+g'^2}{4},\qquad \lambda_3=\frac{g^2-g'^2}{4},\qquad
\lambda_4=-\frac{g^2}{2}, \\
\lambda_5=\lambda_6=\lambda_7=0,\qquad m_{12}^2=m_A^2\cos\beta\sin\beta.
\end{aligned}
\]

The tree-level MSSM is unfortunately not a very interesting model since the mass of the lightest $\cp$-even Higgs boson always fulfills $m_h\leq m_Z$. Such a light $h$ is in contradiction with experiment \cite{Schael:2006cr}, except under special circumstances when the $ZZh$ coupling -- proportional to $\sba$ -- is small. To study the MSSM Higgs sector seriously, it is therefore highly recommended to use one of the specialized tools available \cite{Djouadi:1997yw,*Heinemeyer:1998yj,*Lee:2003nta,*Lee:2007gn}. These include the higher order corrections to the Higgs masses which are necessary to get a phenomenologically viable model. 
In the MSSM at loop level, $\lambda_{5-7}$ will in general be non-zero, and $\cp$ is not necessarily conserved.

\subsection{Example 2: Dark scalar doublet}
\label{sect:dark}
A special case of the 2HDM is when the potential has an exact $Z_2$ symmetry in the Higgs basis, i.e.~in the basis where only one of the two doublets develops a vev \cite{Barbieri:2006dq,Ma:2006km}. In other words, $\lambda_6=\lambda_7=0$, $m_{12}^2=0$, and $\tan\beta=0$.\footnote{Note that this special model cannot be obtained as a continuous $\tan\beta \to 0$ limit from the general potential.} Such a symmetry results in a conserved parity; we call this $H$-parity. 
As a consequence of the conserved quantum number, each scalar mass eigenstate originates from a particular doublet, and no mixing takes place between the doublets. Assigning a positive $H$-parity to the right-handed fermions, couplings to the scalar doublet without a vev (given $H=-1$) are forbidden. The doublet which does not couple to fermions is termed the dark scalar doublet, or sometimes the inert Higgs doublet. Note however that the scalars originating from this doublet still couples to the EW gauge bosons.

The dark scalar doublet model has some interesting phenomenological consequences: First of all, particles with negative $H$-parity have to be produced in pairs, which voids some of the Higgs production mechanisms at the LHC. Second, since the lightest $H$-odd particle is stable, it provides an excellent candidate for scalar dark matter \cite{Ma:2006km}. It has also been argued that this model allows for a heavier ``SM-like'' Higgs boson compatible with the EW precision data \cite{Barbieri:2006dq}.

\section{Constraints}
\label{sect:constraints}

\subsection{Positivity of the potential}
For the vacuum configuration to be stable the Higgs potential, given by eq.~(\ref{eq:pot_gen}), must be positive in all field space directions for asymptotically large values of the fields \cite{Deshpande:1977rw,*Sher:1988mj}. At large field values, the potential is dominated by the quartic terms. These can be conveniently parametrized \cite{ElKaffas:2006nt} by introducing
\begin{equation}
|\Phi_1|=r\cos\gamma, \qquad |\Phi_2|=r\sin\gamma, \qquad  \frac{\Phi_2^\dagger \Phi_1}{|\Phi_1|\,|\Phi_2|} = \rho e^{\rm{i}\theta},
\end{equation}
with $\gamma \in [0,\pi/2]$, $\rho \in [0,1]$, and $\theta \in [0,2\pi)$. Dropping the common factor $r^4$, the quartic terms of the potential thus become
\begin{equation}
  \begin{aligned}
    \mathcal{V}_4=&\frac{1}{2}\lambda_1\cos^4\gamma
    +\frac{1}{2}\lambda_2\sin^4\gamma
    +\lambda_3\cos^2\gamma\sin^2\gamma
    +\lambda_4\rho^2\cos^2\gamma\sin^2\gamma
    \\&+\lambda_5\rho^2\cos^2\gamma\sin^2\gamma\cos(2\theta)
    +\left[\lambda_6\cos^2\gamma
      +\lambda_7\sin^2\gamma
      \right]2\rho\cos\gamma\sin\gamma\cos\theta.
  \end{aligned}
\label{eq:pos}
\end{equation}
Minimizing the expression in eq.~(\ref{eq:pos}), a number of necessary conditions for $\mathcal{V}_4$ to be positive everywhere can be derived:
\begin{equation}
    \lambda_1>0, \qquad \lambda_2>0,\qquad
    \lambda_3>-\sqrt{\lambda_1\lambda_2}.
\label{eq:cond1}
\end{equation}
When $\lambda_6=\lambda_7=0$ the additional condition 
\begin{equation}
\lambda_3+\lambda_4-|\lambda_5|>-\sqrt{\lambda_1\lambda_2},\\
\end{equation}
together with those of eq.~(\ref{eq:cond1}), provide necessary and sufficient conditions for positivity. When either $\lambda_6\neq 0$ or $\lambda_7\neq 0$ the modified condition
\begin{equation}
    \lambda_3+\lambda_4-\lambda_5>-\sqrt{\lambda_1\lambda_2}
\label{eq:cond3}
\end{equation}
applies instead. In addition to the conditions given by eqs.~(\ref{eq:cond1}) and (\ref{eq:cond3}), three more complicated expressions arising from the boundaries $\cos\theta=\pm 1$ and $\rho=1$ are needed in this case. For brevity we have omitted these expressions here, but they are included in the program. 


\subsection{Tree-level unitarity}
In addition to the constraints on the parameters of the potential obtained from positivity, one can also obtain limits by requiring tree-level unitarity for the scattering of Higgs bosons and the longitudinal parts of the EW gauge bosons \cite{Huffel:1980sk,*Maalampi:1991fb,*Kanemura:1993hm,*Akeroyd:2000wc}. 

As is well known, the complete all-order scattering ($S$) matrix has to be unitary. Essentially there are two different ways of achieving this. In a weakly coupled theory higher order contributions to the $S$-matrix typically become smaller and smaller. If the theory on the other hand is strongly coupled, the individual contributions may be arbitrarily large. It is then only in the sum that they cancel and unitarity is respected. 
For a weakly coupled theory it is natural to require that the $S$-matrix is unitary already at tree-level.
Necessary and sufficient conditions on the eigenvalues of the $S$-matrices to achieve this in the general 2HDM have been given in \cite{Ginzburg:2005dt}. Expressed in the parameters $\lambda_i$ of the general potential, the matrices $S_{(Y,\Sigma)}$ for scattering of states with specific total hypercharge $Y$ and weak isospin $\Sigma$, are given by
\begin{equation}
\begin{aligned}
16\pi S_{(2,1)}&=\left(
\begin{array}{ccc}
\lambda_1 & \lambda_5 & \sqrt{2}\lambda_6 \\
\lambda_5 & \lambda_2 & \sqrt{2}\lambda_7 \\
\sqrt{2}\lambda_6 & \sqrt{2}\lambda_7 & \lambda_3+\lambda_4
\end{array}
\right) \\
16\pi S_{(2,0)}&=\lambda_3-\lambda_4\\
16\pi S_{(0,1)}&=\left(
\begin{array}{cccc}
\lambda_1 & \lambda_4 & \lambda_6 & \lambda_6\\
\lambda_4 & \lambda_2 & \lambda_7 & \lambda_7\\
\lambda_6 & \lambda_7 & \lambda_3 & \lambda_5\\
\lambda_6 & \lambda_7 & \lambda_5 & \lambda_3
\end{array}
\right) \\
16\pi S_{(0,0)}&=\left(
\begin{array}{cccc}
3\lambda_1 & 2\lambda_3+\lambda_4 & 3\lambda_6 & 3\lambda_6\\
2\lambda_3+\lambda_4 & 3\lambda_2 & 3\lambda_7 & 3\lambda_7\\
3\lambda_6 & 3\lambda_7 & \lambda_3+2\lambda_4 & 3\lambda_5\\
3\lambda_6 & 3\lambda_7 & 3\lambda_5 & \lambda_3+2\lambda_4
\end{array}
\right). \\
\end{aligned}
\end{equation}
To respect tree-level unitarity, the eigenvalues $\L_i$ of these matrices should fulfill $|\L_i|\leq 16\pi$, which corresponds to saturating the $S$-matrix with the tree-level contribution. Alternatively one can make a partial-wave expansion of the scattering amplitudes for the different channels as discussed in \cite{Gunion:1989we} and put limits on the partial wave amplitudes such as $|\mathrm{Re}(a_j)|<1/2$. The requirement $|\mathrm{Re}(a_0)|<1/2$ for the $J=0$ amplitude then corresponds to $|\L_i|\leq 8\pi$. 
Finally, one can also impose (possibly harder) constraints on the parameters of the potential based on arguments of perturbativity, by demanding that the quartic Higgs couplings fulfill $|C_{H_iH_jH_kH_l}|\leq 4\pi$ or similar constraints.
(In principle one could also imagine a mixed situation where the interaction in some channels is strong and in others perturbative \cite{Ginzburg:2005dt}.)

\subsection{Oblique parameters}
The electroweak oblique parameters $S,T,U$ \cite{Peskin:1990zt,*Altarelli:1990zd,*Peskin:1991sw,*Altarelli:1991fk} constitute a sensitive probe of new physics coupling to the EW gauge bosons. Values on these parameters are constrained by the precision measurements at LEP (see \cite{Amsler:2008zzb}). In general, the contribution to the oblique parameters from the extra $SU(2)$ doublet in the 2HDM is small, since scalar doublets (or singlets) do not break the custodial symmetry \cite{Sikivie:1980hm} which protects the tree-level relation $\rho\equiv M_W/(M_Z\cos\theta_W)=1$. However, large mass splittings among the new Higgs states can still induce sizeable contributions at the loop level.

From the general result of \cite{Grimus:2007if,*Grimus:2008nb}, we determine the 2HDM contribution to the oblique parameters.
For brevity, only the expressions for $S$, $T$ and $U$ are given here, but the full set of six oblique parameters \cite{Maksymyk:1993zm}, {\it i.e}.\ also including $V$, $W$ and $X$, are available in \thdmc. 
Starting with $T$, this parameter is related to the commonly used parameter $\rho_0\equiv\rho/\rho_{\mathrm{SM}}$ through $\rho_0= 1/(1-\alpha T)$. It thus encodes the departure from the SM value of $\rho_0=1$. 
The 2HDM contribution to $T$ is given by
\begin{equation}
\begin{aligned}
T=\frac{g^2}{64\pi^2m_W^2}\Biggl\{&\sum_{k=1}^3\left|C_k\right|^2 F\left(m_{H^+}^2,m_{H_k}^2\right)-\sum_{k=1}^{2}\;\;\left|C_k\right|^2 F\left(m_{H_k}^2,m_A^2\right)\Biggr.\\
&+3\sum_{k=1}^2\left|C_{3-k}\right|^2\Bigl[F\left(m_Z^2,m_{H_k}^2\right)-F\left(m_W^2,m_{H_k}^2\right)\Bigr] \\
&-3\Biggl. \Bigl[F\left(m_Z^2,\mhref^2\right)-F\left(m_W^2,\mhref^2\right)\Bigr] \Biggr\},
\end{aligned}
\label{eq:oblique_T}
\end{equation}
where the masses of the neutral 2HDM Higgs bosons are denoted $m_{H_k}$ ($k=1,2,3$), and $\mhref$ is the mass of the SM Higgs boson. The contribution of the latter is subtracted in order to isolate the new physics contribution. The function $F$ is given by 
\begin{equation}
F(x,y)=\frac{x+y}{2}-\frac{xy}{x-y}\ln\frac{x}{y},
\label{eq:feyn_F}
\end{equation}
and the coupling structure of the EW gauge bosons to a pair of Higgs bosons has been summarized in the vector
\begin{equation}
C=\Bigl\{\cba,\sba,1\Bigr\}.
\end{equation}
Before giving the expressions for the $S$ and $U$ parameters, two additional auxiliary functions, $G$ and $\hat{G}$, are needed. The first of these is defined as 
\begin{equation}
G(x,y,Q)=-\frac{16}{3}+5\frac{x+y}{Q}-2\frac{(x+y)^2}{Q^2}+\frac{3}{Q}\left[\frac{x^2+y^2}{x-y}-\frac{x^2-y^2}{Q}+\frac{(x-y)^3}{3Q^2}\right]\ln\frac{x}{y}+\frac{r}{Q^3}f(t,r),
\label{eq:feyn_G}
\end{equation}
where $t\equiv x+y-Q$, $r\equiv Q^2-2Q(x+y)+(x-y)^2$, and
\begin{equation}
f(t,r)=\left\{
\begin{array}{ll}
\vspace{5pt}
\displaystyle \sqrt{r}\ln\left|\frac{t-\sqrt{r}}{t+\sqrt{r}}\right| & r>0, \\ 
\vspace{5pt}
\displaystyle 0 & r=0,\\
\displaystyle 2\sqrt{-r}\arctan\frac{\sqrt{-r}}{t} & r<0 \;.
\end{array}
\right.
\label{eq:feyn_f}
\end{equation}
The second function $\hat{G}$ is given by
\begin{equation}
\hat{G}(x,Q)=G(x,Q,Q)+12\tilde{G}(x,Q,Q),
\label{eq:feyn_Ghat}
\end{equation}
where
\begin{equation}
\tilde{G}(x,y,Q)=-2+\left[\frac{x-y}{Q}-\frac{x+y}{x-y}\right]\ln\frac{x}{y}+\frac{1}{Q}f(t,r) \;.
\label{eq:feyn_Gtilde}
\end{equation}
Having defined all these functions, the $S$ parameter can then be written simply as
\begin{equation}
\begin{aligned}
S=\frac{g^2}{384\pi^2c_W^2}\Biggl\{&\left[s_W^2-c_W^2\right]^2 G\left(m_{H^+}^2,m_{H^+}^2,m_Z^2\right)+\sum_{k=1}^{2}\left|C_k\right|^2 G\left(m_{H_k}^2,m_A^2,m_Z^2\right)\Biggr. \\
&- 2\ln m_{H^+}^2+\sum_{k=1}^3\ln m_{H_k}^2 - \ln \mhref^2\\
&+\Biggl.\sum_{k=1}^2\left|C_{3-k}\right|^2\hat{G}\left(m_{H_k}^2,m_Z^2\right)-\hat{G}\left(\mhref^2,m_Z^2\right)\Biggr\},
\end{aligned}
\label{eq:oblique_S}
\end{equation}
where $s_W\equiv\sin \theta_W$, $c_W\equiv\cos\theta_W$, and $\theta_W$ is the weak mixing angle. 
Finally, the $U$ parameter is given by
\begin{equation}
\begin{aligned}
U=\frac{g^2}{384\pi^2}\Biggl\{&\sum_{k=1}^3\left|C_k\right|^2 G\left(m_{H^+}^2,m_{H_k}^2,m_W^2\right)-\left[s_W^2-c_W^2\right]^2 G\left(m_{H^+}^2,m_{H^+}^2,m_Z^2\right)\Biggr. \\
&-\sum_{k=1}^{2}\left|C_k\right|^2 G\left(m_{H_k}^2,m_A^2,m_Z^2\right) \\
&+\Biggl.\sum_{k=1}^2\left|C_{3-k}\right|^2\Bigl[\hat{G}\left(m_{H_k}^2,m_W^2\right)-\hat{G}\left(m_{H_k}^2,m_Z^2\right)\Bigr]-\hat{G}\left(\mhref^2,m_W^2\right)+\hat{G}\left(\mhref^2,m_Z^2\right) \Biggr\}.
\end{aligned}
\label{eq:oblique_U}
\end{equation}
The current experimental limits on $S$, $T$ and $U$ can be found in PDG \cite{Amsler:2008zzb}. It should also be noted that typically the $S$ and $T$ parameters cannot be fitted separately from the mass of the SM Higgs boson, $\mhref$. Therefore, when comparing with experimental limits, care has to be taken to use the same value of $\mhref$ when calculating the oblique parameters in the 2HDM at hand, as was done in the SM fit one is comparing to.

\subsection{Anomalous magnetic moment of the muon}
The extended Higgs sector of the 2HDM contributes to the anomalous magnetic moment of the muon, $a_\mu=(g_\mu-2)/2$. Since the experimentally measured value for $a_\mu$ shows a deviation of $\sim 3\,\sigma$ from the SM value \cite{Bennett:2006fi}, this provides an interesting test of new physics models. An overview of the current status, with updated values on $a_\mu$ from both experiment and theory, can be found in \cite{Jegerlehner:2009ry}. 

At one loop, both the charged and neutral Higgs bosons of the 2HDM add to the vertex correction, but these corrections are generally small compared to the observed deviation \cite{PhysRevD.63.091301}. More interesting are the dominant positive contribution at two loops, which is of Barr-Zee type \cite{Barr:1990vd} from diagrams involving a pseudoscalar Higgs, a photon propagator, and a heavy fermion in the upper loop. The large fermion mass compensates with a factor $m_f^2/m_\mu^2$ for the loop suppression. Contributions of similar type, albeit supressed by orders of magnitude, exist also from combinations of $H_iZ$ and $H^\pm W^\pm$. From \cite{Cheung:2003pw,*PhysRevD.64.111301}, we get the 2HDM contribution $\delta a_\mu$ to $a_\mu$ from a photon Barr-Zee diagram with a neutral Higgs and a heavy fermion $f$ in the loop
\begin{equation}
\begin{aligned}
\delta a_\mu^{h} = -&\frac{\alpha}{16\pi^3}N_c Q_f^2 \frac{m_\mu m_f}{m_{h}^2}\Bigl(\kappa^f\sba+\rho^f\cba\Bigr)\Bigl(\kappa^\mu\sba+\rho^\mu\cba\Bigr) f\left(y_f\right)\\
\delta a_\mu^{H} = -&\frac{\alpha}{16\pi^3}N_c Q_f^2 \frac{m_\mu m_f}{m_{H}^2}\Bigl(\kappa^f\cba-\rho^f\sba\Bigr)\Bigl(\kappa^\mu\cba-\rho^\mu\sba\Bigr) f\left(y_f\right)\\
\delta a_\mu^{A} = &\frac{\alpha}{16\pi^3}N_c Q_f^2 \frac{m_\mu m_f}{m_{A}^2}2T_3^f \rho^f\rho^\mu g\left(y_f\right)
\label{eq:barr-zee}
\end{aligned}
\end{equation}
where $y_f=m_f^2/m_{H_i}^2$, $N_c=3(1)$ for quarks (leptons), $Q_f$ the electric charge, and $T_3^f$ the third component of the weak isospin ($T_3=-1/2$ for a down-type fermion). With slight abuse of notation, we have used $\kappa^f,\rho^f$, etc.~to denote the appropriate diagonal elements of the corresponding Yukawa matrices indexed with $U,D,L$ above. Finally, the functions $f$ and $g$ are given by the integrals
\begin{equation}
f(z)=\int_0^1\mathrm{d}x\frac{1-2x(1-x)}{x(1-x)-z}\ln\frac{x(1-x)}{z},
\end{equation}
and
\begin{equation}
g(z)=\int_0^1\mathrm{d}x\frac{1}{x(1-x)-z}\ln\frac{x(1-x)}{z}.
\end{equation}
In \thdmc, the dominant two-loop contributions according to eq.~(\ref{eq:barr-zee}) have been included  for all fermions. For completeness, the sub-dominant contribution from the one-loop corrections has also been included. 

\subsection{Higgs searches at colliders}
\label{sect:lep}
Since no Higgs boson was observed so far either at LEP-II or the Tevatron, stringent limits have been derived on the masses and couplings allowed in the Higgs sector, both within and beyond the SM.
Rather than to implement a full analysis of the collider constraints in the general 2HDM, we provide interfaces for \thdmc\ to use the existing codes \HB\ \cite{Bechtle:2008jh} or \NMSSMT\ \cite{Ellwanger:2004xm,*Ellwanger:2005dv} to check consistency with the existing limits.\footnote{Consult section \ref{sect:install} of this manual for more information on how to install \thdmc\ with \HB\ or \NMSSMT\ support.} \HB\ includes most search channels for neutral Higgs bosons at LEP, and in addition a large number of channels explored by the Tevatron experiments. \NMSSMT\ contains most of the relevant search channels for neutral Higgs bosons at LEP. For an up-to-date list of the channels included, as well as a description of the statistical methods employed to combine the results from several channels, we refer to the original publications \cite{Bechtle:2008jh,Ellwanger:2004xm,*Ellwanger:2005dv} describing these codes.


For the charged Higgs boson, we have implemented the DELPHI limits from the general 2HDM search \cite{Abdallah:2003wd}. The two limits which could be used are both from the pair production process $e^+e^-\to H^+H^-$, followed by decays in the $H^\pm\to\tau^\pm\nu_\tau$ and $H^\pm\to cs$ channels. These two modes are particularly important in the type II 2HDM.

\section{{\tt 2HDMC} code structure}
\label{sect:code}
\subsection{General description}
This section of the manual describes the structure of the \thdmc\ code in terms closely related to the physics. A more technical documentation of the \thdmc\ classes and functions, such as lists of function arguments, are given in the full class documentation distributed with the code.\footnote{\url{http://www.isv.uu.se/thep/MC/2HDMC}}

\thdmc\ is built as a modular {\tt C++} code along standard principles for object-oriented programming. Every program should therefore contain one (or more) \THDM\ object(s) representing the 2HDM(s) under study. Since the 2HDM is an add-on to the SM, each \THDM\ contains its own instance of the \SM\ class. This allows for flexible specification of the underlying SM parameters.

The \THDM\ class itself provides member functions to operate on the 2HDM in various ways, such as changing the basis or the potential parametrization. It further contains the specification of the Yukawa sector, and functions to return the couplings of the 2HDM Higgs bosons. This class also handles input/output to file. To calculate the decay widths and branching ratios of the 2HDM Higgs bosons, the class \DecayTable\ is provided. Objects created from this class will operate on a \THDM. The same is true for the \Constraints\ class, which contains functions to calculate the various observables discussed in section~\ref{sect:constraints}.

\subsection{Physics input}
\begin{table}
\centering        
\begin{tabular}{cccc}
   \toprule          
         Parameter  & Value & Parameter & Value  \\
         \midrule
         $\alpha^{-1}(m_Z)$   & $127.91$ & $G_F$ & $1.16637\times 10^{-5}$ GeV$^{-2}$ \\
         $\alpha_s(m_Z)$      & $0.1176$ & $m_Z$ & $91.1876$ GeV \\
         \bottomrule
\end{tabular}
 \caption{Default values used in \thdmc\ for the Standard Model gauge parameters.}
\label{tab:SM}
\end{table}
\begin{table}
\centering
\begin{tabular}{cccccc}
   \toprule
         Parameter  & Value & Parameter & Value & Parameter & Value \\
        \midrule
         $m_u$   & $0$      & $m_d$ & $0$     & $m_e$     & $0.511\times 10^{-3}$ \\
         $m_c$   & $1.77$   & $m_s$ & $0.1$   & $m_\mu$   & $0.10566$\\
         $m_t$   & $171.2$  & $m_b$ & $4.8$   & $m_\tau$  & $1.777$\\
        \bottomrule
\end{tabular}
 \caption{Default values used in \thdmc\ for the fermion pole masses. All values are in GeV.}
\label{tab:qmass}
\end{table}
The Standard Model parameters in \thdmc\ are contained in objects of the \SM\ class. Default values for all parameters (taken from PDG \cite{Amsler:2008zzb}) are stored as static members in this class. The default values are tabulated in table~\ref{tab:SM} (for the gauge sector parameters) and table~\ref{tab:qmass} (fermion masses). To obtain the \SM\ object which contains the SM parameters for an instance of \THDM, the function {\tt get\_SM} should be used. The parameters can then be changed by invoking different functions, \emph{e.g.}~to set the top quark mass {\tt set\_qmass} can be used. For a complete list we refer to the \SM\ class documentation. Finally, to use the new parameters, the function {\tt set\_SM} is called for the \THDM\ with the modified \SM\ object as argument. An example of how to set $m_t$ is given in the {\tt Demo} program. 

In the EW sector, the SM is specified by three parameters: $\alpha$, $G_F$, and $m_Z$, from which $m_W$ and the weak mixing angle $\sin^2\theta_W=1-m_W^2/m_Z^2$ are determined. From these basic parameters, values for the supplementary EW quantities are computed according to
\begin{equation}
v = (\sqrt{2}G_F)^{-1/2},\qquad g = \frac{2m_W}{v}, \qquad g' = \frac{2m_W}{v}\tan\theta_W, \qquad e = \frac{2m_W}{v}\sin\theta_W.
\end{equation}

The \thdmc\ user can enter the Higgs potential parameters in a number of different parametrizations using member functions of \THDM. A full list of the available parametrizations, and their corresponding parameters, is given in table~\ref{tab:potinput}. For each function {\tt set\_param\_X} listed in this table, there is a corresponding function {\tt get\_param\_X} which can be used to retrieve the current model in that parametrization.

\begin{table}
\centering        
\begin{tabular}{lll}
   \toprule          
         \THDM\ function  & Required parameters & Comment \\
         \midrule
         {\tt set\_param\_gen} & $\lambda_1,\,\lambda_2,\,\lambda_3,\,\lambda_4,\,\lambda_5,\,\lambda_6,\,\lambda_7,\,m_{12}^2,\,\tan\beta$ & General potential, eq.~(\ref{eq:pot_gen}) \\
         {\tt set\_param\_higgs} & $\Lambda_1,\,\Lambda_2,\,\Lambda_3,\,\Lambda_4,\,\Lambda_5,\,\Lambda_6,\,\Lambda_7,\,m_{H^\pm}$ & Higgs basis, $\tan\beta=0$\\
         {\tt set\_param\_hhg} & $\Lambda_1,\,\Lambda_2,\,\Lambda_3,\,\Lambda_4,\,\Lambda_5,\,\Lambda_6,\,\tan\beta$ & $Z_2$-symmetry only softly broken \\
         {\tt set\_param\_phys} & $m_h,\,m_H,\,m_A,\,m_{H^\pm},\,\sba,\,\lambda_6,\,\lambda_7,\,m_{12}^2,\,\tan\beta$ & Physical masses \\

        \bottomrule
\end{tabular}
 \caption{Input parameters for the Higgs potential in different bases and parametrizations.}
\label{tab:potinput}
\end{table}

A few additional comments about table~\ref{tab:potinput} are in place. The number of required input parameters depends on the assumptions underlying the specific parametrization. For instance, in the Higgs basis $\tan\beta\equiv 0$ and need not be specified. Another example is given by the potential in the Higgs Hunter's Guide \cite{Gunion:1989we} which was originally introduced in \cite{Georgi:1978xz}. This parametrization only allows for soft breaking of the $Z_2$ symmetry, equivalent to forcing $\lambda_6=\lambda_7=0$ in the general potential. Note the different meaning of $\Lambda_i$ in this parametrization compared to the general case (the relation between the two can be found, for example, in Appendix A of \cite{Gunion:2002zf}).
In the physical basis, the Higgs masses are given as input together with the invariant $\sba$. The remaining parameters in this parametrization are of course not more physical than in any general potential. Since $\lambda_6,\lambda_7$, and $m_{12}^2$ play a role in the Higgs self-interactions, their effects appear for instance in the loop-mediated decay $H\to\gamma\gamma$. In the conventions used, the allowed range for $\alpha$ is fixed by the requirements $0 \leq \beta \leq \pi/2$, and $-\pi/2 \leq \beta-\alpha \leq \pi/2$.

Having specified the Higgs potential, we can move on to specify the Yukawa sector in order to have a complete model. This is only required when Higgs couplings to fermions are going to be considered. Several \THDM\ functions, listed in table~\ref{tab:yukinput}, exist for specifying the Yukawa sector according to eq.~(\ref{eq:yukawa}). Non-diagonal elements in the $\rho$ matrices are allowed, but will lead to FCNCs. There are presently no checks against limits on FCNC included in the program. It is important to note that $\kappa$ is defined internally using pole masses. When related to $\kappa$, the elements of $\rho$ should therefore also be specified using pole masses. To resum large logarithms, the flavour-diagonal couplings $H_iq\bar{q}$ are rescaled with $\overline{m}_q(m_{H_i})/m_{q}^{\mathrm{pole}}$. Specifying $\rho$ matrices directly related to e.g.~the $\overline{\mathrm{MS}}$ masses can therefore lead to unwanted consequences, such as a broken $Z_2$ symmetry in the Yukawa sector.

\begin{table}
\centering        
\begin{tabular}{lll}
   \toprule                
         \THDM\ function  & Arguments & Comment \\
         \midrule
         {\tt set\_yukawas\_type} & Type ($1$--$4$) & Universal Yukawa type (see table~\ref{tab:types}) \\
         {\tt set\_yukawas\_up} & $\rho^U_{uu}$, $\rho^U_{cc}$, $\rho^U_{tt}$ & Elements of diagonal $\rho^U$ matrix\\
         {\tt set\_yukawas\_up} & $\rho^U_{ij}$ & Elements of full $\rho^U$ matrix \\        \bottomrule
\end{tabular}
 \caption{Functions which are used to specify the up-type quark Yukawa sector. Similar methods exist for the down-type quark and lepton Yukawas. Note that {\tt set\_yukawas\_type} specifies all sectors simultaneously. Yukawa matrices related to quark masses should always be specified in terms of the quark pole masses.}
\label{tab:yukinput}
\end{table}

Two special cases of the 2HDM have been implemented as special input functions in \THDM. These are described in table~\ref{tab:specinput}. What distinguishes these are that they set both the potential parameters and the Yukawa sector using a single function. Not surprisingly, {\tt set\_MSSM} sets the 2HDM parameters according to the tree-level MSSM relations (see section \ref{sect:MSSM}), with the Yukawa sector of the MSSM (type II). Again we emphasize that the tree-level MSSM in only included for reference, and that it is not intended to be used in any physics analysis. The other special function is {\tt set\_inert}, which sets the parameters of the dark scalar doublet model (section \ref{sect:dark}). The convention chosen for the parameters of this model is that of \cite{Gustafsson:2007pc,*Lundstrom:2008ai}. Only one doublet is given Yukawa couplings in this case.
\begin{table}
\centering        
\begin{tabular}{lll}
   \toprule                
         \THDM\ function  & Required parameters & Comment \\
         \midrule
         {\tt set\_MSSM} & $m_A,\,\tan\beta$ & Tree-level MSSM potential, Yukawas type II \\
         {\tt set\_inert} & $m_h^{\mathrm{SM}},\,m_H^{\mathrm{DS}},\,m_A^{\mathrm{DS}},\,m_{H^+}^{\mathrm{DS}},\,\lambda_2,\,\lambda_3$ & Dark scalar (DS) doublet has no vev, no Yukawas\\
        \bottomrule
\end{tabular}
 \caption{Special functions used for giving input parameters to the Higgs potential, including specification of the Yukawa sector.}
\label{tab:specinput}
\end{table}

\subsection{Program output}
The main output which is possible to get from \thdmc\ can be summarized as follows:
\begin{itemize}
\item Potential parameters in different parametrizations and bases
\item Higgs masses and couplings
\item Higgs decay widths and branching ratios
\item Properties of interest for constraining the 2HDM parameter space
\end{itemize}
It is very simple to use \thdmc\ for converting between different parametrizations of the Higgs potential. Simply give the parameters using one of the {\tt set\_param} functions, followed by a {\tt get\_param} to get another representation of the same model. The basis can also be changed in this way. Invoking {\tt recalc\_tan\_beta}, the model parameters are recalculated for a new value of $\tan\beta$. Of course this does only work for the $U(2)$-invariant Higgs potential, not for the Yukawa sector. To write out the parameters in a certain model to {\tt stdout}, a number of \THDM\ functions starting with {\tt print\_param} are available.

When the Higgs masses are not used as input, they are calculated from the tree-level Higgs potential. The mass values can be accessed either through {\tt get\_param\_phys}, or using {\tt get\_hmass} which takes an index for the Higgs boson as argument. For every method where this is the case, we use the enumeration $\left\{h=1,H=2,A=3,H^\pm=4\right\}$. The couplings involving one or more Higgs bosons are available through the functions listed in table~\ref{tab:couplings}.
\begin{table}
\centering
\begin{tabular}{llll}
\toprule
\THDM\ function & Coupling & Feynman rule & Convention \\
\midrule
{\tt get\_coupling\_hdd} & $H_i\,\bar{d}_jd_k $ & $C_S+\gamma_5 C_P$\\
{\tt get\_coupling\_huu} & $H_i\,\bar{u}_ju_k$ &$C_S+\gamma_5 C_P$& \\
{\tt get\_coupling\_hll} & $H_i\,\bar{\ell}_j\ell_k$ &$C_S+\gamma_5 C_P$& \\
{\tt get\_coupling\_hud} & $H^+\bar{d}_ju_k$ &$C_S+\gamma_5 C_P$& \\
{\tt get\_coupling\_hln} & $H^+\ell^+_j\nu_{k}$ &$C_S+\gamma_5 C_P$& \\
{\tt get\_coupling\_vvh} & $H_i V_j V_k$ & $Cg^{\mu\nu}$& $C$  in GeV\\
{\tt get\_coupling\_vhh} & $H_i H_j V_k$ & $C(p_i-p_j)^\mu$& All momenta ingoing\\
{\tt get\_coupling\_hhh} & $H_i H_j H_k$ & $C$& $C$  in GeV\\
{\tt get\_coupling\_vvhh} & $H_i H_j V_k V_l$ &$Cg^{\mu\nu}$& Single $W^+$ momenta ingoing\\
{\tt get\_coupling\_hhhh} & $H_i H_j H_k H_l$& $C$& \\
\bottomrule
\end{tabular}
\caption{Functions which are used to calculate the Higgs boson couplings (denoted $C$ in the table). For a purely neutral current interaction, $i=1\dots 3$. When both neutral and charged currents are possible, $i=1\dots 4$ with $H^+$ as $i=4$. Vector indices go from $1$ to $3$, corresponding to $\gamma$, $Z$, and $W^+$. Fermion family indices go from $1$ to $3$. The flavour-diagonal couplings to quarks assume renormalization of quark $\overline{\mathrm{MS}}$ masses at the Higgs mass scale.}
\label{tab:couplings}
\end{table}

To calculate the partial decay widths of the 2HDM Higgs bosons into various channels, the  class \DecayTable\ is used. The functions available through this class are collected in table~\ref{tab:gammas}. To toggle QCD corrections and loop-induced decay modes ($H_i\to gg,\gamma\gamma$) off/on, the method {\tt set\_QCD} can be used.
\begin{table}
\centering
\begin{tabular}{lll}
\toprule
\DecayTable\ function & Width & Comments \\
\midrule
{\tt get\_gamma\_hdd} & $\Gamma(H_i\to\bar{d}_jd_k)$ & NLO QCD corr. \\
{\tt get\_gamma\_huu} & $\Gamma(H_i\to\bar{u}_ju_k)$ & NLO QCD corr. \\
{\tt get\_gamma\_hll} & $\Gamma(H_i\to\bar{\ell}_j\ell_k)$ & \\
{\tt get\_gamma\_hud} & $\Gamma(H^+\to\bar{d}_ju_k)$ & LO QCD corr.\\
{\tt get\_gamma\_hln} & $\Gamma(H^+\to\ell^+_j\nu_{k})$ & \\
{\tt get\_gamma\_hvv} & $\Gamma(H_i\to V_j^{(*)} V_j)$ &  \\
{\tt get\_gamma\_hvh} & $\Gamma(H_i\to V_j^{(*)} H_k)$ & Widths for $W^+ H^-$ and $W^-H^+$ added\\
{\tt get\_gamma\_hhh} & $\Gamma(H_i\to H_j H_k)$ & \\
{\tt get\_gamma\_hgaga} & $\Gamma(H_i\to \gamma\gamma)$ & Effective operator\\
{\tt get\_gamma\_hgg} & $\Gamma(H_i\to gg)$ & Effective operator, LO QCD corr.\\
\bottomrule
\end{tabular}
\caption{Functions which calculate the decay widths of Higgs boson $H_i$, $i=1\dots 4$. Fermion family indices go from $1$ to $3$, whereas gauge boson indices go from $2$ to $3$, meaning in turn $Z$ and $W^\pm$. The decay widths for $H \to V^{(*)}V$ and $H \to V^{(*)}H$ also include the decay through a virtual gauge boson.}
\label{tab:gammas}
\end{table}

As discussed in section~\ref{sect:constraints}, there exist theoretical constraints on the 2HDM parameters. Functions to check that these are satisfied are provided in the \Constraints\ class. This class also provides the calculation of the 2HDM contribution to the other observables discussed in section~\ref{sect:constraints}. Table~\ref{tab:constraints} lists the various functions that can be of interest when using \thdmc\ to constrain the 2HDM parameters. 

Another set of 2HDM constraints is given by various flavour-physics observables, such as rare decays of B-mesons. No flavour observables are included in \thdmc, but instead the code can be easily interfaced with SuperIso \cite{Mahmoudi:2007vz,*Mahmoudi:2008tp}.
\begin{table}
\centering
\begin{tabular}{llll}
\toprule
\Constraints\ function & Description & Returns \\
\midrule
{\tt check\_positivity} & Checks potential positivity & {\tt true/false}\\
{\tt check\_unitarity} & Checks tree-level unitarity & {\tt true/false}\\
{\tt check\_perturbativity} & Checks perturbativity & {\tt true/false}\\
{\tt check\_masses} & Checks the collider constraints & {\tt true/false}\\
{\tt oblique\_param} & Calculates EW oblique parameters ($S,T,U,V,W,X$) & values\\
{\tt delta\_amu} & Calculates the 2HDM contribution to $a_\mu$ & value \\
\bottomrule
\end{tabular}
\caption{Functions which are available in the \Constraints\ class. Functions with names beginning with {\tt check\_} give boolean {\tt true} for models that comply with the required condition, and {\tt false} otherwise. The other functions return the 2HDM contributions for comparison to experimental limits.}
\label{tab:constraints}
\end{table}

\subsection{LesHouches Interface}
The input to \thdmc\ can also be specified through a file interface inspired by the LesHouches format \cite{Skands:2003cj,*Allanach:2008qq}. To distinguish it from parameter files for the MSSM, the file must contain
\begin{verbatim}
Block MODSEL
   0   10  
\end{verbatim}
to select the 2HDM. The Higgs potential input and basis choice can be given in three formats, corresponding to the general basis, the Higgs basis, and physical basis as described above. For the general and Higgs bases, the relevant blocks and parameter numbers that need to be specified are given in table~\ref{tab:lhgen}, while those for the physical basis are given in table~\ref{tab:lhphys}.

When $\tan\beta\neq 0$, the type of Yukawa couplings can also be specified using {\tt MINPAR(24)} to select between types $1$--$4$. The LesHouches interface  cannot be used to specify more generic Yukawa couplings at this point. A sample input file named {\tt LH\_input} is included in the \thdmc\ distribution. This file can be run with the example program {\tt CalcLH}.
\thdmc\ can also produce LesHouches-formatted output by calling the \THDM\ function
{\tt write\_LesHouches} with the filename specified. For an example of how such an output file might look, see appendix~\ref{sect:LHout}.
\begin{table}
\centering
\begin{tabular}{lll}
\toprule
Generic basis &  & \\
\midrule
{\tt MINPAR} &  &\\
{\tt 3} & $\tan\beta$ & Set to zero for Higgs basis \\
{\tt 11-17} & $\lambda_1$--$\lambda_7$ & \\
{\tt 18} & $m_{12}^2$  & Generic basis only\\
{\tt MASS} & & \\
{\tt 37} & $m_{H^+}$ & Higgs basis only \\
\bottomrule
\end{tabular}
\caption{LesHouches input parameters for the generic and Higgs bases}
\label{tab:lhgen}
\end{table}
\begin{table}
\centering
\begin{tabular}{lll}
\toprule
Physical basis & & \\
\midrule
{\tt MINPAR} & &\\
{\tt 3} & $\tan\beta$ &\\
{\tt 16}--$17$ & $\lambda_6$--$\lambda_7$ &\\
{\tt 18} & $m_{12}^2$ &\\
{\tt 20} & $\sba$ &\\
{\tt MASS} & & \\
{\tt 25} & $m_h$ &Lightest $\cp$-even Higgs \\
{\tt 35} & $m_H$ &Heaviest $\cp$-even Higgs \\
{\tt 36} & $m_A$ &$\cp$-odd Higgs \\
{\tt 37} & $m_{H^+}$ &Charged Higgs \\
\bottomrule
\end{tabular}
\caption{LesHouches input parameters for the physical mass basis.}
\label{tab:lhphys}
\end{table}

\subsection{MadGraph/MadEvent model}
A {\tt USRMOD} model for \MGME\ \cite{Alwall:2007st} is provided with \thdmc\ to facilitate event generation in the 2HDM. In the {\tt MGME} subdirectory of the \thdmc\ distribution, a directory called \thdmc\ contains the model. This directory should be placed in the {\tt Models} directory of \MGME. Using the standard output function {\tt write\_LesHouches} with the {\tt couplings} argument produces model output in the format of a {\tt  param\_card}. All couplings listed in table~\ref{tab:couplings}, except the four-Higgs vertices, are included in the \thdmc\ model. At this point the model cannot handle flavour-changing neutral currents. To use tree-level widths consistently, the \thdmc\ QCD corrections to Higgs decay widths into fermions and the decays $H_i\to gg,\gamma\gamma$ can be swtiched off using the {\tt set\_QCD} function of \DecayTable.

One advantage with our model over the ``default'' 2HDM model in \MGME\ \cite{Alwall:2007st} is the treatment of couplings and decay widths. All input to our model is computed with \thdmc\ to ensure internal consistency. The $H_i \bar{q}q$ couplings (evaluated at the correct Higgs mass scale) are then input to the model. With the default model, these couplings are computed in the model itself, whereas the decay widths are computed externally by a calculator. An obvious advantage with the existing model is that it allows for simulation of processes that violate $\cp$, which our model does not. To use the output of \thdmc\ together with the default 2HDM model should require only minor modifications of the LesHouches file format used.

\section{Downloading and compiling the package}
\label{sect:install}
The latest version of the \thdmc\ code can be obtained from
\begin{verbatim}
    http://www.isv.uu.se/thep/MC/2HDMC.
\end{verbatim}
In addition, this manual and the complete class documentation, including the specification of the required input and output of various functions, can be downloaded from the same URL. 
To compile \thdmc, the following tools are required:
\begin{itemize}
\item A {\tt C++} compiler, such as {\tt GCC}
\item Common build utilities: {\tt make} etc.
\item The GNU Scientific Library ({\tt GSL}); available from \url{http://www.gnu.org/software/gsl/}
\end{itemize}

The \thdmc\ library has been successfully compiled using {\tt GCC 4} under {\tt Linux}.  However, there should be no difficulties in using {\tt GCC 3} or another compiler and/or architecture. Please report to the authors your experiences from compiling and running \thdmc\ on another platform.
Unpacking the main tarball, a directory called {\tt 2hdmc} is created. To start the compilation, enter this directory and run
\begin{verbatim}
   $ make
\end{verbatim}
to compile the library and some example programs (file names starting with {\tt Calc}) that demonstrate the use of \thdmc\ in standalone applications. The installation can be verified by running the {\tt Demo} program and comparing the output to appendix~\ref{app:output}. To compile the \thdmc\ library only, run
\begin{verbatim}
   $ make lib
\end{verbatim}
while a main program ({\tt Program.cpp}, located in {\tt 2hdmc/src}) is compiled with the \thdmc\ library using the command
\begin{verbatim}
   $ make Program
\end{verbatim}
We encourage the user to examine the {\tt Makefile} for details.

As described in section~\ref{sect:lep}, there are two possibilities to access constraints from collider experiments in \thdmc . These can be used either independently or in combination, and neither is required to run \thdmc. However, if you are using any of these codes with your \thdmc\ project, make sure to cite the relevant publication.

\HB \footnote{\url{http://www.ippp.dur.ac.uk/HiggsBounds}} \cite{Bechtle:2008jh} provides limits from both LEP and the Tevatron experiments. To use \HB, the library {\tt libHB.a} should be placed in the {\tt 2hdmc/lib} directory, and in addition two lines should be uncommented in the \thdmc\ {\tt Makefile}. The interface has been tested to run with \HB\ 1.0.3.

\NMSSMT \footnote{\url{http://www.th.u-psud.fr/NMHDECAY/nmssmtools.html}}  \cite{Ellwanger:2004xm,*Ellwanger:2005dv} includes limts from LEP only. To use \NMSSMT, the library {\tt nmhdecay.a} and the experimental data provided in the distribution is needed. The library should be put in the {\tt 2hdmc/lib} directory, and the experimental data must be accessible in {\tt ../EXPCON} when running a program compiled to use \NMSSMT. To compile \thdmc\ with \NMSSMT\ support, three lines should be uncommented in the {\tt Makefile}. Version 2.2.0 has been tested to work.

\bibliographystyle{JHEP}
\bibliography{manual}
\newpage
\appendix
\section{Sample output}
\label{app:output}
\thdmc\ ships with numerous example main programs. Compiling and running {\tt Demo.cpp} 
should produce the following output to {\tt stdout}:\\
\begin{centering}
\rule{\textwidth}{0.01cm}
\end{centering}
\begin{verbatim}
$ ./Demo
****************************************************
*                                                  *
*    2HDMC - Two-Higgs-Doublet Model Calculator    *
*        http://www.isv.uu.se/thep/MC/2HDMC        *
*                  Version 1.0.3                   *
*             Compiled on Jun 22 2009              *
*                                                  *
****************************************************

2HDM parameters in physical mass basis:
      m_h:     80.00000
      m_H:    200.00000
      m_A:    140.00000
     m_H+:    160.00000
 sin(b-a):      0.20000
 lambda_6:      0.00000
 lambda_7:      0.00000
    m12^2:    800.00000
tan(beta):     10.00000

2HDM parameters in generic basis:
 lambda_1:      2.25553
 lambda_2:      0.61585
 lambda_3:      2.29636
 lambda_4:     -0.38796
 lambda_5:     -0.19002
 lambda_6:      0.00000
 lambda_7:      0.00000
    m12^2:    800.00000
tan(beta):     10.00000

2HDM parameters in Higgs basis:
 Lambda_1:      0.63763
 Lambda_2:      2.24484
 Lambda_3:      2.29082
 Lambda_4:     -0.39350
 Lambda_5:     -0.19556
 Lambda_6:     -0.10861
 Lambda_7:     -0.05374
     m_Hp:    160.00000

Constraints:
  Potential stability: OK
 Tree-level unitarity: OK
       Perturbativity: OK
     Mass constraints: OK

 Oblique S:  6.77200e-02
 Oblique T: -1.62816e-02
 Oblique U:  2.95734e-04
 Oblique V:  7.70150e-03
 Oblique W:  1.02287e-03
 Oblique X:  7.70645e-04
 Delta_rho: -1.27289e-04

 Delta_amu:  3.78746e-11

 Total width for h :  1.707e-01 GeV
 Total width for H :  5.212e+00 GeV
 Total width for A :  2.979e-01 GeV
 Total width for H+:  4.083e-02 GeV
\end{verbatim}

\section{LesHouches output}
\label{sect:LHout}
The same {\tt Demo} program discussed in appendix~\ref{app:output} produces the LesHouches file
{\tt Demo\_out.lha} with the following content:\\
\begin{centering}
\rule{\textwidth}{0.01cm}
\end{centering}
\begin{verbatim}
##################################################################
#                                                                #
#          Two-Higgs Doublet Model LesHouches-card               #
#       Produced by 2HDMC: www.isv.uu.se/thep/MC/2hdmc           #
#                                                                #
##################################################################
Block MODSEL # Select Model
    0   10     #  10 = THDM
Block SMINPUTS  # Standard Model inputs
    1        1.27910000e+02   # 1/alpha_em(MZ) SM MSbar
    2        1.16637000e-05   # G Fermi
    3        1.17600000e-01   # alpha_s(MZ) SM MSbar
    4        9.11876000e+01   # MZ
    5        4.24693805e+00   # mb(mb)
    6        1.75000000e+02   # mt (pole)
    7        1.77700000e+00   # mtau(pole)
Block MINPAR    # Model parameters
# Parameters for general potential in generic basis
    3        1.00000000e+01   # tan(beta)
   11        2.25553058e+00   # lambda_1
   12        6.15852536e-01   # lambda_2
   13        2.29636045e+00   # lambda_3
   14       -3.87961523e-01   # lambda_4
   15       -1.90021971e-01   # lambda_5
   16        0.00000000e+00   # lambda_6
   17        0.00000000e+00   # lambda_7
   18        8.00000000e+02   # m_12^2
   20        2.00000000e-01   # sin(beta-alpha)
   21        9.79795897e-01   # cos(beta-alpha)
   24     2                   # Yukawas Type
Block MGCKM     # CKM elements
    1     1        9.74190000e-01   # Vud
    1     2        2.25700000e-01   # Vus
    1     3        3.59000000e-03   # Vub
    2     1        2.25600000e-01   # Vcd
    2     2        9.73340000e-01   # Vcs
    2     3        4.15000000e-02   # Vcb
    3     1        8.74000000e-03   # Vtd
    3     2        4.07000000e-02   # Vts
    3     3        9.99133000e-01   # Vtb
Block MASS      #  Mass spectrum (kinematic masses)
#  PDG      Mass
     1        0.00000000e+00   # Md
     2        0.00000000e+00   # Mu
     3        1.00000000e-01   # Ms
     4        1.77000000e+00   # Mc
     5        4.80000000e+00   # Mb
     6        1.75000000e+02   # Mt
    11        5.10998918e-04   # Me
    13        1.05658367e-01   # Mmu
    15        1.77700000e+00   # Mta
    23        9.11876000e+01   # MZ
    24        7.98257288e+01   # MW
    25        8.00000000e+01   # Mh1, lightest CP-even Higgs
    35        2.00000000e+02   # Mh2, heaviest CP-even Higgs
    36        1.40000000e+02   # Mh3, CP-odd Higgs
    37        1.60000000e+02   # Mhc
Block alpha     # Effective Higgs mixing parameter
              1.26976975e+00   # alpha
#     PDG   Width
DECAY  6     1.55688294e+00   # top decays
#            BR          NDA    ID1   ID2
       9.98881018e-01     2      24     5
       2.25882632e-08     2      37     1
       4.91600304e-07     2      37     3
       1.11846795e-03     2      37     5
DECAY  25     1.70669488e-01   # h1 decays, lightest CP-even Higgs
#            BR          NDA    ID1   ID2
       1.06791622e-03     2       3    -3
       6.75570041e-05     2       4    -4
       9.06817540e-01     2       5    -5
       7.40019764e-09     2      11   -11
       3.16378237e-04     2      13   -13
       8.92259685e-02     2      15   -15
       3.28604940e-06     2      22    22
       2.50134614e-03     2      21    21
DECAY  35     5.21246842e+00   # h2 decays, heaviest CP-even Higgs
#            BR          NDA    ID1   ID2
       8.17946511e-06     2       3    -3
       4.80480461e-05     2       4    -4
       6.08736565e-03     2       5    -5
       5.83863108e-11     2      11   -11
       2.49619285e-06     2      13   -13
       7.05732616e-04     2      15   -15
       2.39195048e-05     2      22    22
       6.81763412e-02     2      23    23
       1.94337426e-01     2      24   -24
       1.72829055e-04     2      21    21
       7.30433309e-01     2      25    25
       2.59594580e-06     2      23    36
       8.78581628e-07     2      24   -37
       8.78581628e-07     2      37   -24
DECAY  36     2.97857573e-01   # h3 decays, CP-odd Higgs
#            BR          NDA    ID1   ID2
       1.14103374e-03     2       3    -3
       6.83455111e-06     2       4    -4
       8.99212561e-01     2       5    -5
       8.05509800e-09     2      11   -11
       3.44380190e-04     2      13   -13
       9.73790965e-02     2      15   -15
       3.62240562e-07     2      22    22
       1.07345981e-03     2      21    21
       8.42263714e-04     2      23    25
DECAY  37     4.08333142e-02   # Charged Higgs decays
#            BR          NDA    ID1   ID2
       4.69922214e-04     2       2    -3
       9.17812917e-05     2       2    -5
       2.75348241e-06     2       4    -1
       8.78870607e-03     2       4    -3
       1.22618690e-02     2       4    -5
       6.71516661e-08     2     -11    12
       2.87094082e-03     2     -13    14
       8.11866349e-01     2     -15    16
       1.63559670e-01     2      24    25
       8.79405283e-05     2      24    36
#
\end{verbatim}

\end{document}